\documentclass[]{article}

\usepackage{amsmath,amssymb}
\usepackage{bm}
\usepackage{booktabs}
\usepackage{graphicx}
\usepackage{subcaption}
\usepackage[table]{xcolor}
\usepackage{colortbl}
\RequirePackage{hyperref}
\usepackage{lmodern}

\begin{document}

\title{Analysing Collective Behaviour in Temporal Networks Using Event Graphs and Temporal Motifs}
\author{Andrew Mellor\footnote{mellor@maths.ox.ac.uk} \\ \small Mathematical Institute, University of Oxford, Oxford, OX2 6GG, UK}
\date{}
\maketitle

\begin{abstract}
  Historically studies of behaviour on networks have focused on the behaviour of individuals (node-based) or on the aggregate behaviour of the entire network.
  We propose a new method to decompose a temporal network into macroscale components and to analyse the behaviour of these components, or collectives of nodes, across time.
  This method utilises all available information in the temporal network (i.e. no temporal aggregation), combining both topological and temporal structure using temporal motifs and inter-event times.
  This allows us create an embedding of a temporal network in order to describe behaviour over time and at different timescales.
  We illustrate this method using an example of digital communication data collected from an online social network.
\end{abstract}

\section{Introduction}

Modern methods of data collection have given researchers access to ever richer datasets, both in terms of their descriptive power and their temporal resolution.
Where previously data would be a representation of an instant in time or aggregated over a time interval, in many fields we now have access to sequential (and often timestamped) records or observations.
The mining of this sequential data, or \emph{event streams}, is now an important task with a particular focus being on creating algorithms that extract information in real-time.

Event stream data is perhaps most prevalent in the digital context, where data collected can be easily automated.
Online social networks such as Twitter \cite{kwak2010twitter} allow access to real-time streams of messages between users of the platform. 
Similarly, researchers and telecom providers have access to telephone and text records 
(call participants, durations, etc.) which can be analysed to describe our circadian rhythms \cite{monsivais2017tracking}, closest friends \cite{onnela2007structure}, or the differences in call behaviour between genders \cite{kovanen2013temporal}.
Event stream data is not limited to communication however.
Other examples include, but are limited to, website clickstreams \cite{banerjee2001clickstream}, proximity networks \cite{barrat2013temporal}, transportation networks \cite{pan2011path}, and biological interactions \cite{han2004evidence}.  

To address the challenges that come with temporal data there have been many recent advances in the study of temporal networks \cite{masuda2016guide, holme2013temporal, holme2015modern}.
Temporal networks are a natural extension from static networks although there are many ways to model them depending on the degree of temporal aggregation and the type of analysis required.  
Typically temporal networks are aggregated into time intervals (binning events into intervals of an hour for example) to create a sequence of static networks which can then be analysed using concepts adapted from static graph theory \cite{grindrod2014dynamical,holme2015modern}.
However, arbitrarily discretising time can distort results by destroying temporal correlations and obscuring causal relations.
More recent work has avoided this issue by considering a network of temporal events (time-stamped edges) as opposed to individual nodes themselves \cite{mellor2017temporal, kivela2017mapping}.
This allows the temporal network to be represented as a single weighted static network which can be decomposed by pruning edges based on edge weights. 
The connected components of this network provide a natural partition of the data whereby events in close temporal and topological proximity remain connected.

In this article we describe a new method to analyse collective behaviour in temporal networks using the network decomposition in \cite{mellor2017temporal, kivela2017mapping} and cluster the resulting temporal components using a combination of temporal and topological features.
In Section~\ref{sec:methodology} we detail our methodology and feature section.
In particular we introduce three new features in the form of the motif entropy, inter-event time entropy, and the degree imbalance.
Sections~\ref{sec:data} and \ref{sec:results} are devoted to the application of the method to data collected from Twitter, and finally in Section~\ref{sec:conclusions} we discuss the challenges and issues with such methods and highlight avenues for further research. 

\subsection{Related Work}

This article combines and extends higher-order network models, temporal motifs, and network embedding.

Motifs have been previously used to understand behaviour in networks \cite{jurgens2012temporal, kovanen2013temporal, kovanen2011temporal}. 
In \cite{kovanen2013temporal} it is found that there are differences in the call behaviour of each gender (measured by the differences in temporal motif distributions), although the change of this behaviour over time is not investigated.
The behaviour of Wikipedia editors is characterised in \cite{jurgens2012temporal} using temporal motifs, who show that the observed behavioural patterns are relatively stable over the course of a number of years.

Network (or node) embedding involves characterising a network (or node in a network) by an n-dimensional \emph{feature vector}.
After embedding, methods that have traditionally been used for image-recognition or natural language processing such a deep neural networks or random forests can be used for link prediction, clustering, or graph synthesis.
There have numerous suggestions for network (or node) embeddings \cite{hamilton2017representation, grover2016node2vec, henderson2012rolx, perozzi2014deepwalk}.
Node embeddings can be generated by considering local sampling of the network around the node, either iteratively \cite{henderson2012rolx} or through a random walk exploration \cite{grover2016node2vec, perozzi2014deepwalk}.
Walk-based embeddings are well grounded (having connections to the graph Laplacian), although the embeddings themselves can be difficult to interpret.
Similar approaches have been implemented to characterise subgraphs \cite{duvenaud2015convolutional} or entire graph structures \cite{ikehara2017characterizing}.
The latter instead using a set of predefined features (clustering, assortativity, etc.) to create an embedding, rather than a random walk exploration.

More recently there have been attempts to extend node embeddings to the dynamic case.
One approach creates node embeddings by considering random \emph{temporal walks} \cite{nguyen2018continuous}.
These are random walks that have the added constraint of traversing edges in the order at which they appear.
There are approaches to embed dynamic graphs \cite{rossi2012time} however these are reliant on discretising time and creating a sequence of aggregate static networks.

The representation of the temporal network we use is a second-order time-unfolded model, a combination of the higher-order models of \cite{scholtes2017network} and time-unfolded networks of \cite{michail2016introduction}. 
The decomposition of temporal network consists of using the connected components these higher-order models (more detail given in Section~\ref{sec:methodology}). 
This is the first instance of embedding and clustering these components, so there is currently not a directly comparable benchmark.
We instead compare our decomposition to the more typical decomposition {fixed-width} time intervals. 

The example in Sections~\ref{sec:data} and \ref{sec:results} considers directed messages on Twitter.
Many studies of Twitter use the content of messages (text and media) to model topics of conversation, perform sentiment analysis, and identify viral content \cite{bollen2011twitter,sakaki2010earthquake,ramage2010characterizing,panisson2014mining}.
Topics are generated by aggregating messages which have similar word usage.
We instead take the content-agnostic approach and instead model only the connections between users.

\section{Methodology}
\label{sec:methodology}

In this section we describe the methods used to model and analyse temporal network data.
Our analysis is split into three parts; temporal decomposition, feature extraction, and clustering.

\subsection{Event Graphs and Temporal Decomposition}

Formally we describe a temporal network $G$ by the triple $(V,T,E)$ where $V \subset \mathbb{N}$ is a set of nodes, $T \subset \mathbb{R}_0^+$ a set of interaction times, and $E \subset V^2 \times T$, the set of temporal events.
Directed, instantaneous events take the form $e_i = (u_i,v_i,t_i)$ corresponding to an edge appearing from node $u_i$ to node $v_i$ at time $t_i$.
The remainder of this article we assume that events are both directed and instantaneous, although the methods can easily be generalised to undirected or persistent events.

We model the temporal network as an event graph (EG) \cite{mellor2017temporal}. 
An EG is a second-order time-unfolded model of the temporal network, where the events of the original network are the nodes of the EG.
The EG is described by the tuple $\mathcal{G} = (E,L)$, where $E$ is the set of events, and $L$ is the set of edges.
Events are linked to the subsequent events for each node in that event (resulting in a directed acyclic graph with a maximal in/out degree of two).
Weighting these edges with the time between the two events (inter-event time) gives rise to a weighted, directed graph of temporal events.
Each edge is also associated with a two-event temporal motif \cite{kovanen2011temporal} which describes the topological relationship between the nodes in each event (see Figure~\ref{fig:motifs}).

The {$\Delta t$-EG} is a subgraph of the EG where the edges are removed based on the edge weights.
The new set of edges is $L_{\Delta t} = \{(e_i, e_j) \in L | t_j-t_i \le \Delta t\}$ with the {$\Delta t$-EG} given by $\mathcal{G}_{\Delta t} = (E, L_{\Delta t})$.
This allows us to study the temporal-topological structure of the {$\Delta t$-EG} as a function of $\Delta t$, and decompose the event sequence into smaller subsequences.
Let $C_{\Delta t}$ be the set of weakly-connected components of the $\Delta t$-EG.
Each component $c \in C_{\Delta t}$ is described by the tuple $(E^c, L_{\Delta t}^c)$ with $\bigcup_c E^c = E$ and $\bigcup_c L_{\Delta t}^c = L_{\Delta t}$.
The temporal components (which we will use synonymously with weakly-connected component) of $\mathcal{G}_{\Delta t}$ capture sequences of events which occur in relatively close proximity and share one or more nodes which we assume to imply a level of causality between events.

Algorithmic details of how to construct the EG are given in Appendix~\ref{app:sec:eg_algorithm}.

\subsection{Feature Extraction and Component Embedding}

How can we characterise these temporal components?
We consider a set of features which can be split into purely temporal, purely topological, and topological-temporal features.
For this study we will consider only features which are independent of the temporal component size, or \emph{scale-invariant}.
Here the component `size' refers to number of events, the number of nodes, or the duration of the component.
Naturally this allows us to compare behaviour in components of different sizes and across different datasets.

In the rest of this section we describe the features used in this study.
A full list of features considered is given in Table~\ref{table:feature_list}.

\subsubsection{Topological-Temporal Features}

\begin{figure}[!h]
	\centering
	\begin{subfigure}{0.12\linewidth}
		\centering
		\includegraphics[width=\linewidth]{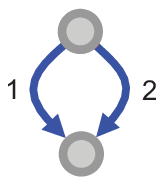}
		\caption*{ABAB}
	\end{subfigure}
	\begin{subfigure}{0.12\linewidth}
		\centering
		\includegraphics[width=\linewidth]{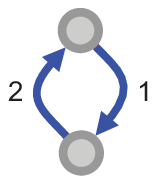}
		\caption*{ABBA}
	\end{subfigure}
	\begin{subfigure}{0.12\linewidth}
		\centering
		\includegraphics[width=\linewidth]{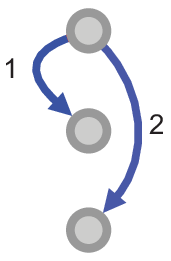}
		\caption*{ABAC}
	\end{subfigure}
	\begin{subfigure}{0.12\linewidth}
		\centering
		\includegraphics[width=\linewidth]{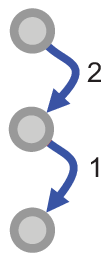}
		\caption*{ABCA}
	\end{subfigure}
	\begin{subfigure}{0.12\linewidth}
		\centering
		\includegraphics[width=\linewidth]{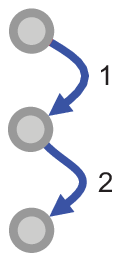}
		\caption*{ABBC}
	\end{subfigure}
	\begin{subfigure}{0.12\linewidth}
		\centering
		\includegraphics[width=\linewidth]{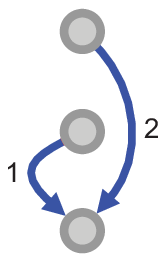}
		\caption*{ABCB}
	\end{subfigure}
	\caption{
		All possible two-event temporal motifs.
		Events are labelled with the order which they occur.
	}
	\label{fig:motifs}
\end{figure}

\textbf{Temporal motifs.} 
Temporal motifs are repeatedly observed patterns of interaction across time.
We consider temporal motifs as defined in \cite{kovanen2011temporal} although we can equally use the definition in \cite{paranjape2017motifs}\footnote{
  The two definitions differ in two aspects.
  Motifs in \cite{kovanen2011temporal} require events for each node in the motif to be consecutive and the time between consecutive events has an upper bound whereas events in \cite{paranjape2017motifs} do not need to be consecutive and the entire motif must occur within a fixed time window (with no prescription of inter-event time). 
  } 
by considering a different event graph structure \cite{mellor2018event}.
In particular we consider only two-event motifs, of which there are six (shown in Figure~\ref{fig:motifs} along with a shorthand description of each).
We can also incorporate different event types (or event colours) into a temporal motif (see Appendix~\ref{app:sec:temporal_motifs}).
For example, if there are two event types x and y the possible motifs derived from the ABBA motif are ABxBAx, ABxABy, AByBAx, and AByBAy.
There are 24 possible motifs of this variety $(6 \times 2^2)$.
Modelling multiple event types allows us to better distinguish between different types of behaviour in temporal networks (the differences between calls and text in a communication network, for example), however it increases the number of possible motifs.
We measure the prevalence of each motif type present in a component by the fraction of all observed motifs.
This gives us an empirical motif distribution for the component.
Typically the prevalence of motifs is uninformative without reference to a baseline model \cite{kovanen2011temporal,kovanen2013temporal}, or a comparative dataset.
In this case since we are comparing different temporal components (and not commenting on relative prevalence) we can use the motif counts as they are.
We do however compare our method to an ensemble of time-shuffled replicates of our data in Section~\ref{sec:results}.

We are also interested in how diverse a component is in terms of the motifs present.
To quantify this we introduce the entropy of the motif distribution, i.e.
\begin{align}
  S_{\rm MOT} = \sum_{m \in M} p_m \log_2 (p_m)
\end{align}
where $M$ is the set of all possible motifs, and $p_m$ is the probability of observing motif type $m$.
This has the desired property that $S_{\rm MOT} = 0$ when only one type of motif is present, and takes a maximal value of $\log_2{|M|}$ when all $|M|$ motif types are equally likely.
This suggests the rescaling 
\begin{align}
  \hat{S}_{\rm MOT} = \frac{S_{\rm MOT}}{\log_2{|M|}}
\end{align}
which maps to the unit interval.

Temporal motifs have been previously used to analyse behaviour in temporal networks and so should be a suitable feature to distinguish behaviour across the constituent temporal components of the network.
We could also consider higher-order motifs (containing three or more events) however the number of possible motifs increases exponentially with the motif size.
This makes these higher-order motifs both costly to compute and also poorer features on small networks as the counts for each motif are likely to be small.

\subsubsection{Pure Topological Features}

These features consider the static graph aggregation of the temporal component given by the adjacency matrix
\begin{align*}
	A^{(c)}_{uv} = \begin{cases}
		1 & \text{ if } \exists t \text{ s.t } (u,v,t) \in E^c \\
		0 & \text{ otherwise.}
	\end{cases}
\end{align*}
That is, an edge is present in the aggregated graph if an event occurred along that edge in the duration of the temporal component.
While this removes any temporal information about the component it allows us to consider higher-order properties of the network without explicitly calculating higher-order temporal motifs.
The structural features of static graphs have been well documented \cite{newman2010networks,costa2007characterization}.
Here we consider three well known features, and introduce a new feature, the \emph{degree imbalance}.

\textbf{Clustering coefficient.} 
The clustering coefficient $C$ measures the degree of transitivity in the graph
and is given by the ratio number of closed triplets of nodes to the number of connected triplets of nodes.
This gives a measure of the likelihood that nodes $a$ and $b$ are connected, given a connection between $a$ and $c$ and between  $c$ and $b$ also.
For this feature we ignore the direction of edges.

\textbf{Degree Imbalance.} 
The degree imbalance measures the average difference between the degrees of connected nodes.
Let $\alpha,\beta \in \{{\rm in, out }\}$ index the degree type, and let $s_i^\alpha$ and $t_i^\beta$ be the $\alpha$- and $\beta$-degree of the source and target node for edge $i$.
The degree imbalance is given by
\begin{align}
  \mu_{(\alpha,\beta)} = \frac{ m^{-1} \sum_i (s^\alpha_i - t^\beta_i)}{\max_i |s^\alpha_i - t^\beta_i|}
\end{align}
where $m$ is the number of edges in the graph, and we sum over all possible edges.
We also normalise by the maximum difference between node degrees.
The degree imbalance takes values in $[-1,+1]$ with a value of $\mu_{(\alpha,\beta)} = \pm 1$ indicating 
that for all edges the $\alpha$-degree of the source is greater/less (resp.) than the $\beta$-degree of the target, and this difference is the same for all edges.
By considering the joint degree distribution it can be shown that $\mu{\rm(in,out)}=0$ for all graphs and so this case will be omitted.

We use this feature to assess how `hub-like' the network is.
This feature is used, as opposed to the directed degree assortativity \cite{foster2010edge} as the degree assortativity is not well defined on networks where there is no variance in the node degrees (and this case is not addressed). 
An example of such a case is an inward star graph where the out-degree of all edge sources (the peripheral nodes) is one, and the in degree of all edge targets (the hub) is the number of peripheral nodes.

Normalisation to the unit interval can be achieved with
\begin{align}
  \hat{\mu}_{(\alpha,\beta)} = \frac{\mu_{(\alpha,\beta)} + 1}{2}.
\end{align}

\textbf{Edge reciprocity.} 
The edge reciprocity is a measure of how many events in the conversation are reciprocated and is given by
\begin{align}
	R = m^{-1}\sum_{i,j} A_{ij}A_{ji}
\end{align}
where $m$ is the total number of edges as before.
This feature picks up instances of reciprocation that are not captured in an ABBA motif (either due to other intermediate motifs being formed, or the two events not being close in temporal proximity).

\textbf{Edge density.} 
The edge density is the number of edges present in the graph as a fraction of all possible edges. Mathematically this is given by
\begin{align}
	\rho = \frac{1}{N(N-1)}\sum_{i,j} A_{ij},
\end{align}
where $N$ is the number of nodes in the graph.
The presence of an $N^2$ term in the denominator indicates that this feature is only scale-independent on the assumption that the network is dense (i.e. node degrees $k_i$ are $\mathcal{O}(N)$).
Since we do not wish to make that assumption, the edge density will only be used as a descriptor and not a feature in the model.

\subsubsection{Pure Temporal Features}

\textbf{Inter-event times (IETs).}
The times between consecutive events has been the focus of many studies \cite{goh2008burstiness,lambiotte2013burstiness,stehle2010dynamical}.
We consider the distribution of IETs along all edges of the EG, that is the times between consecutive events for each node in the temporal component.
The IET distribution is dependent on the duration of the conversation (the maximal IET is bounded by the conversation duration) and so cannot be used as a scale-invariant feature.
We instead characterise the diversity of the IETs by considering
the entropy, as with motifs.
The IET is a continuous variable however so we first need to discretise it into fixed-width bins.
Let $I$ be the set of intervals that partition the set of IETs and $p_i$ be the probability of an IET being in interval $i \in I$, then the IET entropy is given by 
\begin{align}
  S_{\rm IET} = -\sum_{i \in I} p_i \log_2 (p_i).
\end{align}
The IET entropy has the property of being zero for periodic events and is maximal for a uniform distribution of IETs.
The entropy is desirable over the variance in the case of events with alternating periods, e.g. $2$-$4$-$2$-$4$.
In this example, the variance is dependent on the distance between the two periods and can be arbitrarily large whereas the entropy is constant and is close to zero regardless of distance between the two periods.
We normalise the IET entropy by dividing by the maximum possible entropy for that number of bins, i.e. 
\begin{align}
  \hat{S}_{\rm IET} =  \frac{S_{\rm IET}}{\log_2{|I|}}.
\end{align}

\textbf{Activity.}
Another related temporal feature we consider is the activity of the temporal component, denoted by $\lambda$, and defined to be the number of events per unit time.
To normalise the activity feature we apply the transformation
\begin{align}
\hat{\lambda} = 1-e^{-\lambda} 
\end{align}
which takes values in $[0,1)$.

\subsubsection{Summary statistics}

The summary features we will use to describe temporal components are
the \emph{number of events}, the \emph{number of nodes}, and the total duration or \emph{lifetime} of the component.
For a component consisting of event set $E$ these are
\begin{align}
N_{\rm events}(E) &= |E|,\\
N_{\rm nodes}(E) &= \left| \bigcup_{(u,v,t) \in E} (u \cup v) \right|,\\
D(E) &= \max_{(u,v,t) \in E} t - \min_{(u,v,t) \in E} t,
\end{align}
respectively.
As these features by definition explicitly depend on the size of the component we will not include them in our feature-space.
They do however provide important context to examine individual components.

\begin{table}[!h]
  \centering
  \begin{tabular}{@{}lllc@{}} \toprule[1pt]
  \textbf{Feature}  & \textbf{Symbol} & \textbf{Range} & \textbf{Scale-invariant} \\ \midrule[0.5pt]
\# Nodes &  $N_{\rm nodes}$    & $[2,\infty)$ &  No \\
\# Events &  $N_{\rm events}$   & $[2,\infty)$  &  No \\ 
Duration (seconds) &  $D$    & $[1,\infty)$ & No \\
Edge Density &  $\rho$             & $[0,1]$ &   No  \\\midrule[0.1pt]
Motif Prevalence (multiple) & $p_m$ & $[0,1]$ & Yes \\
Motif Entropy &  $S_{\rm MOT}$     & $[0,\log_2{|M|}]$ &   Yes    \\
IET Entropy &  $S_{\rm IET}$       & $[0,\log_2{|I|}]$ &   Yes    \\
Degree Imbalance (in,in) &  $\mu_{({\rm in,in})}$    & $[-1,1]$ &  Yes   \\
Degree Imbalance (out,in) &  $\mu_{({\rm out,in})}$   & $[-1,1]$ &  Yes   \\
Degree Imbalance (out,out) &  $\mu_{({\rm out,out})}$    & $[-1,1]$ &  Yes    \\
Clustering &  $C$           & $[0,1]$ &   Yes    \\
Reciprocity &  $R$           & $[0,1]$ &   Yes    \\
Activity &  $\lambda$       & $[0,\infty)$ & Yes   \\ \bottomrule[1pt]
  \end{tabular}
  \caption{
    List of features used to describe temporal components.
    Features are rescaled so that all values lie in $[0,1]$ (see text).
    The prevalence of each motif is an individual feature, meaning for two-event motifs with two possible event types there are 24 individual features. 
  }
  \label{table:feature_list}
\end{table}

\subsection{Embedding \& Clustering}

We create a feature vector $\bm{x}_c$ for each temporal component using the above features.
Formally we define an encoding function $f: C_{\Delta t} \to \mathbb{R}^d$ which maps a temporal component into the $d$-dimensional vector space.
In total we have $32$ different features, or dimensions.
We then normalise our feature vectors by rescaling to unit length, i.e. $\bm{\hat{x}}_c = \bm{x}_c / |\bm{x}_c|$ .

To cluster temporal components we first need to define a suitable distance function.
There are many ways to do this however we choose to use the Euclidean distance between temporal components.
This gives us pairwise distance matrix between all temporal components.
From the distance matrix we can construct a hierarchical clustering of components by employing Ward's method \cite{ward1963hierarchical}.
Finally we use the silhouette coefficient  \cite{rousseeuw1987silhouettes} to determine the optimal number of clusters.
The silhouette coefficient score for a sample $i$ is given by
\begin{align}
  s(i) = \frac{d_{\rm inter}(i)-d_{\rm intra}(i)}{\max(d_{\rm intra}(i),d_{\rm inter}(i))}
\end{align} 
where $d_{\rm intra}$ and $d_{\rm inter}$ are the intra-cluster distance (distance to centre of assigned cluster) and inter-cluster distance (distance to the centre of the nearest alternate cluster) respectively.
The silhouette coefficient for the dataset is then the average across all samples and takes values in $[-1,1]$ with a value of $+1$ indicating perfect clustering and a value of $-1$ indicating the worst possible clustering.

\section{Data}
\label{sec:data}

There are many sources of timestamped event data, such as electronic communication, website clickstreams, proximity networks, or transportation schedules, for example.
To illustrate our method we consider data collected from Twitter and investigate whether we can identify different types of distinct collective behaviours in the conversations that users have online.

We take Twitter data sampled by keyword using the Twitter Streaming API\footnote{
   See \href{developer.twitter.com}{developer.twitter.com}
   }. 
In particular we collect tweets which contain the word \emph{Emirates} (and common variants).
We subsequently collate the accounts present in the sample and collect all tweets they have participated in during the course of the entire day using the REST API.
This procedure is important for two reasons.
Firstly, we want to be able to capture all conversations surrounding a particular topic and subsequent responses to sampled tweets may or may not mention the keyword explicitly.
Secondly, as the temporal motifs are formed of \emph{subsequent} events, the omission of any event will introduce error.

\begin{table}[!h]
  \centering
  \begin{tabular}{@{}cccc@{}} \toprule[1pt]
  \textbf{\# Tweets} & \textbf{\# Events} & \textbf{\# Users } &  \textbf{\% Retweets} \\ \midrule[0.5pt]
  161,730 & 130,360 & 48,249 & 52.95 \\ \bottomrule[1pt]
  \end{tabular}
  \caption{
    Data collected from Twitter over a $24$ hour period.
    Tweets with no mentions (i.e. are not directed) do not constitute an event hence there are fewer events than tweets.
    }
  \label{table:dataset}
\end{table}

We sampled our keyword using the above procedure over one day in November 2017.
The statistics of the data sampled are presented in Table~\ref{table:dataset}. 

\subsection{Modelling Considerations}

In the data collection we collect two types tweet, \emph{messages} (original pieces of content) and \emph{retweets} (duplicates of previous tweets which are reshared).
We model these tweets as sequence of temporal events $E$ where each event $e_i \in E$ takes the form $(u_i,v_i,t_i)$ which describes the instantaneous contact from node $u_i$ to node $v_i$ at time $t_i$.
Here a tweet from account $u$ to account $v$ (directing their message to them) at time $t$ becomes a temporal event $(u, v, t)$.
If multiple accounts are mentioned within a tweet then multiple events are created\footnote{
  In this study we restrict ourselves to events containing only two accounts/nodes.
  We can however extend this analysis to multiple-node events (or \emph{hyper-events}).
  }.
A retweet by account $u$ of a tweet from account $v$ at time $t$ becomes an event $(v, u, t)$ which reflects the direction that information has travelled in the event.
These two types of event can be considered as either a \emph{push} (message) or \emph{pull} (retweet) of information.

\subsection{Time-shuffled Ensembles}

Without a suitable reference model we instead benchmark our findings using ensembles of randomised time-shuffled data.
To shuffle the data we permute the event times randomly while keeping the node pairs constant.
This preserves network structure and the numbers of events between all pairs of nodes but destroys all temporal correlation between events \cite{gauvin2018randomized}. 

\section{Results}
\label{sec:results}

We construct the EG and set the edge (inter-event time) threshold to be {$\Delta t = 240$s} (or four minutes).
The 4min-EG has 4137 components with at least five events.
This choice of $\Delta t$ was chosen to balance the number of temporal components with component size; too small a value would possibly split up conversations and create components of only a small number of events, and too large a value leads to fewer large components that may be unrelated.
Further discussion of this choice is given in Appendix~\ref{app:sec:deltat}.
In Figure~\ref{fig:barcode} we plot the activity of the ten largest components in an hour period.
Immediately we can see the differences of the temporal activity patterns of these components; some components occur over the space of 10 minutes (C10), others span the entire interval (C4).
We can also see clear distinctions between the inter-event times for the components, for example the contrast between the bursty behaviour of component C6 and component C5 where activity levels peak and then decay.

\begin{figure}[!h]
	\centering
	\includegraphics[width=0.8\linewidth]{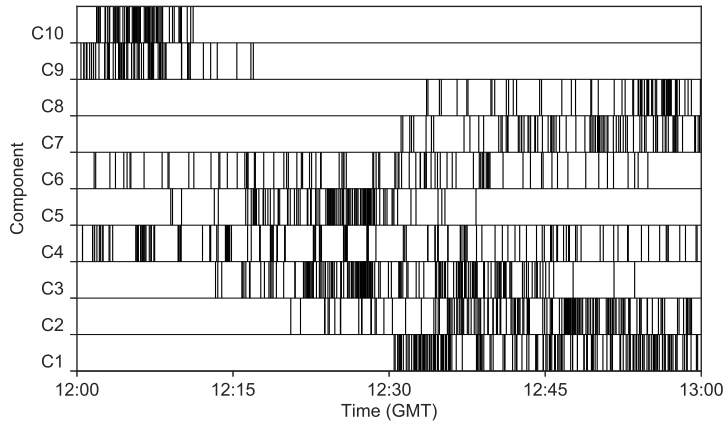}
	\caption{
		Temporal barcode for a one hour  window of the {4min-EG}.
		The largest 10 temporal components (by number of events) of the {4min-EG} are plotted (C1 being the largest).
		Each vertical black line represents the time at which an event occurred.
		We see distinct patterns in behaviour in terms of event density, component duration, and inter-event time distributions (gaps between events).
	}
	\label{fig:barcode}
\end{figure}

\subsection{Discriminative features}

\begin{figure}[!h]
\centering
\includegraphics[width=0.8\linewidth]{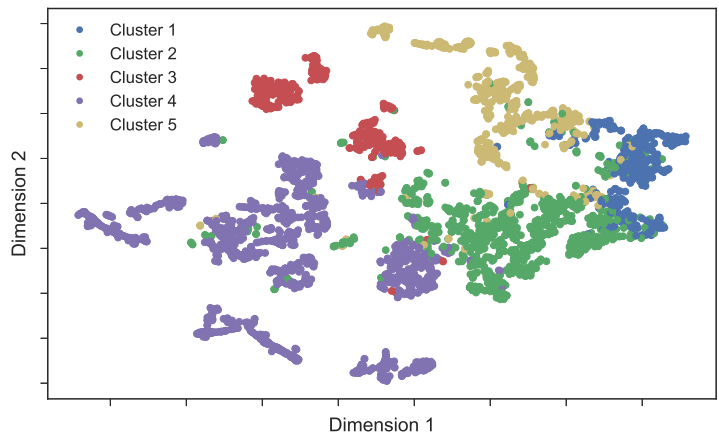}
  \caption{
  Scatterplot of the 4137 component features, using the t-SNE algorithm \cite{maaten2008visualizing} to reduce feature-space to two dimensions.
  The cluster analysis is conducted on the full feature space and dimensionality reduction is used solely for visualisation.
  Components are coloured by their assigned clustering (see text).
  }
  \label{fig:clustering}
\end{figure}

The component barcodes portray only the duration, activity, inter-event times, and number of events in each component.
We can systematically investigate which are the most important features by applying principal component analysis (PCA) to reduce the dimensionality of the feature-space.
The first three components of the PCA account for 71\% of the variance of component features.
Visualising the temporal components in the full feature-space is not feasible and so we instead reduce the dimensionality of the data using the t-SNE algorithm \cite{maaten2008visualizing} (Figure~\ref{fig:clustering}).

The high-dimensionality makes the results difficult to plot but does not hamper the rest of our analysis which is conducted on the full feature-space.
\begin{table}[!h]
  \centering
  \textbf{PCA Component} \\
  \begin{tabular}{@{}rrr@{}} \toprule[1pt]
   \textbf{1} (36\%)     & \textbf{2} (19\%)    &  \textbf{3} (15\%)  \\ \midrule[0.5pt]
   ABrCBr & ABmCBm     & $S_{\rm IET}$    \\
   $\mu_{\rm (out,in)} $      & $\mu_{\rm (out,in)} $  & ABrABr    \\
   ABmCBm    & $S_{\rm MOT}$      & $R$    \\ \bottomrule[1pt]
  \end{tabular}
    \caption{The top three feature contributions to the components of the PCA (ranked by magnitude).}
  \label{table:components}
\end{table}

In Table~\ref{table:components} we show the largest constituent features of the components of the PCA, ordered by magnitude.
The motif entropy is also one of the distinguishing features - how diverse the conversation is in terms of the motifs present.
This is enlightening as low motif entropy can be indicative of predictable, systematic behaviour, possibly highlighting automated accounts.
Similarly, low IET entropy implies periodic (or predictable) activity, which again provides further evidence of automation. 
The presence of reciprocity is also distinguishing feature of an online conversation, i.e., is whether the conversation is `one-sided', or just consists of back-and-forth interaction.
Degree imbalance being a prominent feature of the PCA also suggests that whether or not the conversation is `hub-like' is another way to distinguish conversations.

\subsection{Clustering}

The temporal components are partitioned into five clusters shown by the dendrogram in Figure~\ref{fig:dendrogram}.
We choose five clusters in particular as this maximises the silhouette score (see Appendix~\ref{app:sec:clusters}).
\begin{table}[!h]
  \centering
  \begin{tabular}{@{}lrrrrr@{}} \toprule[1pt]
    & \multicolumn{5}{c}{\textbf{Conversation Cluster}} \\
  \textbf{Feature}  & \textbf{1} & \textbf{2} & \textbf{3} & \textbf{4} &\textbf{5} \\ \midrule[0.5pt]

  $N_{\rm nodes}$*    &       5.33 &      22.46 &       8.15 &      14.49 &      13.98 \\
  $N_{\rm events}$*   &      13.67 &      33.11 &      12.13 &      20.21 &      16.76 \\
  $D$ (seconds)*      &     542 &     686 &     166 &     455 &     751 \\
  $\rho$*         &       0.42 &       0.16 &       0.18 &       0.13 &       0.15 \\\midrule[0.1pt]
  ABAB Motif        &       0.09 &        \cellcolor{orange}{0.12} &       \cellcolor{orange}{0.34} &        \cellcolor{orange}{0.19} &        \cellcolor{orange}{0.16} \\
  ABAC Motif        &       0.07 &       0.05 &       \cellcolor{red}{0.40} &       \cellcolor{yellow}{0.02} &       \cellcolor{yellow}{0.08} \\
  ABBA Motif        &       \cellcolor{orange}{0.28} &       0.11 &       0.00 &       0.00 &       0.01 \\
  ABBC Motif        &       0.12 &       0.08 &       0.02 &       0.00 &       0.01 \\
  ABCA Motif        &       \cellcolor{yellow}{0.16} &       \cellcolor{yellow}{0.11} &       0.01 &       0.00 &       0.02 \\
  ABCB Motif        &       \cellcolor{red}{0.29} &       \cellcolor{red}{0.53} &       \cellcolor{yellow}{0.23} &       \cellcolor{red}{0.79} &       \cellcolor{red}{0.71} \\ \midrule[0.1pt]
  $S_{\rm MOT}$     &       1.85 &       1.74 &       1.33 &       0.48 &       0.68 \\
  $S_{\rm IET}$        &       1.92 &       1.97 &       0.98 &       1.81 &       2.04 \\
  $\mu({\rm in,in})$    &      -0.27 &      -0.55 &      -0.79 &      -0.97 &      -0.88 \\
  $\mu({\rm out,in})$    &       0.08 &      -0.59 &       0.42 &      -0.93 &      -0.62 \\
  $\mu({\rm out,out})$     &       0.27 &       0.11 &       0.83 &       0.91 &       0.80 \\
  $C$             &       0.41 &       0.02 &       0.02 &       0.00 &       0.02 \\
  $R$           &       0.48 &       0.18 &       0.00 &       0.00 &       0.03 \\
  $\lambda$       &       0.03 &       0.03 &       0.57 &       0.06 &       0.03 \\ \bottomrule[1pt]
  \multicolumn{6}{r}{*not used as a feature}
  \end{tabular}
  \caption{Feature vector averages across clusters (pre-normalisation). 
  For brevity, we have combined the motifs with different event types.
  The first, second, and third most prevalent motif in each cluster is highlighted in red, orange, and yellow respectively.}
  \label{table:component_statistics}
\end{table}

\begin{figure}[!h]
	\centering
	\begin{subfigure}{0.3\linewidth}
		\centering
		\includegraphics[width=\linewidth]{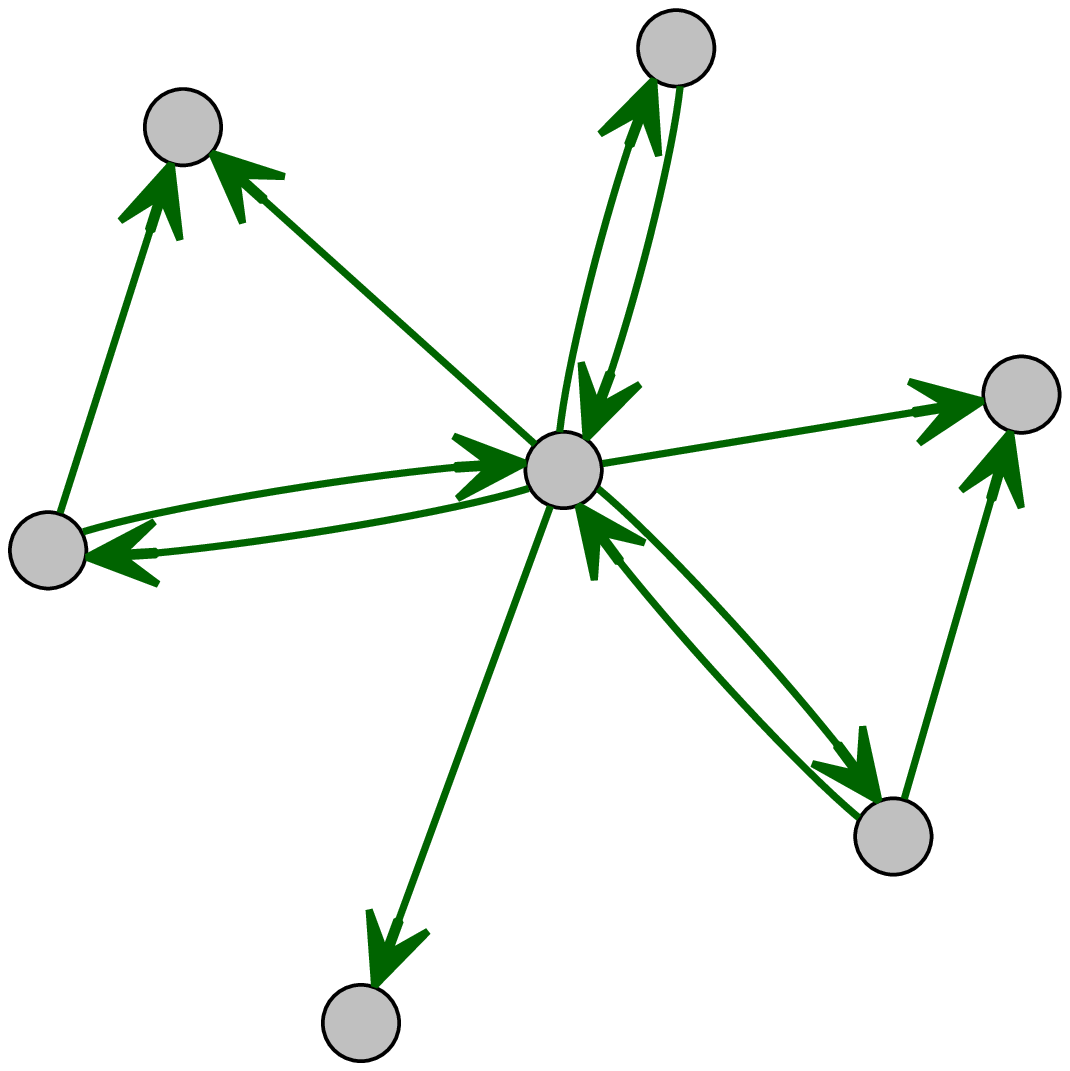}
		\caption{Cluster 1}
	\end{subfigure}
	\begin{subfigure}{0.3\linewidth}
		\centering
		\includegraphics[width=\linewidth]{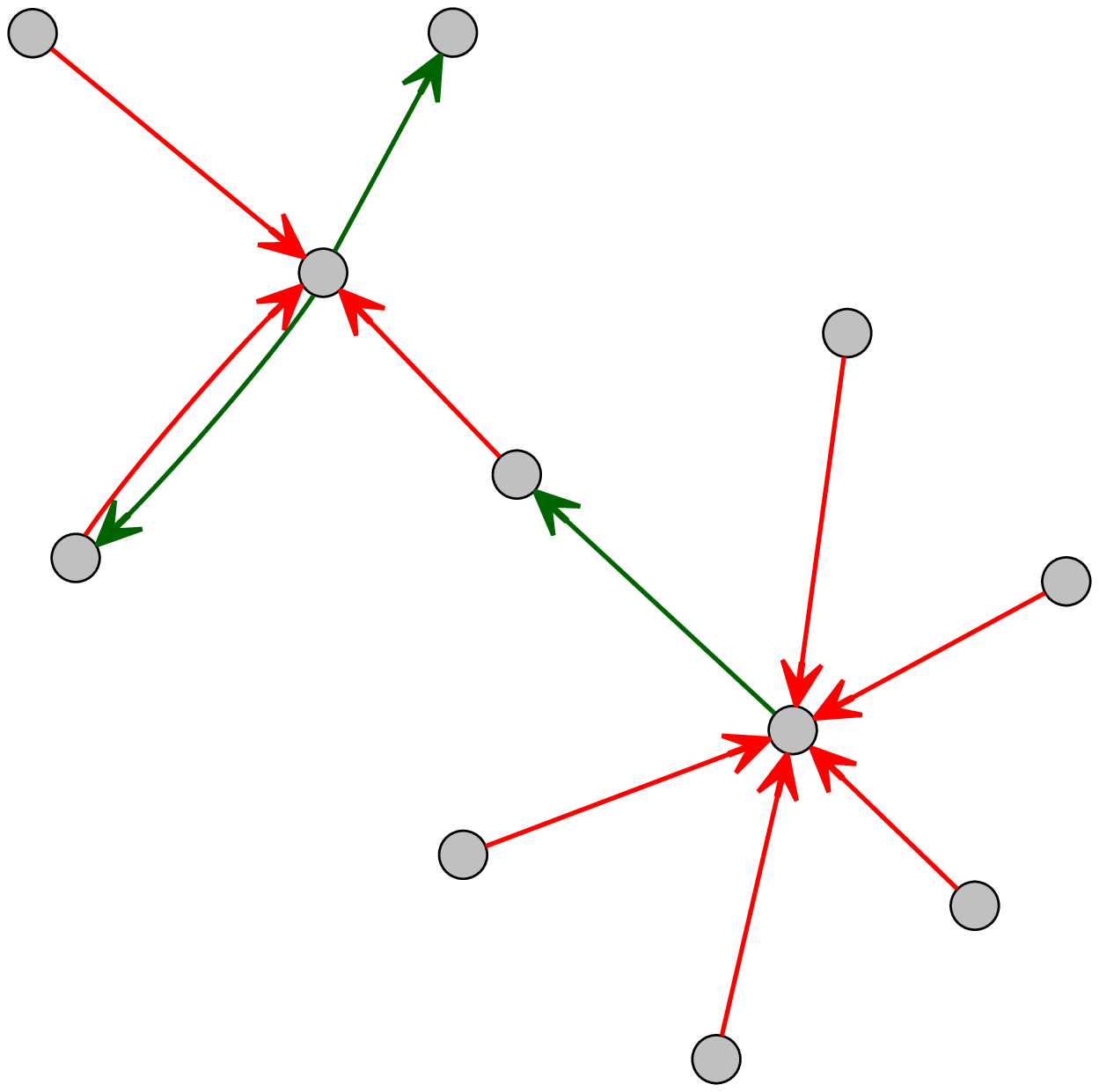}
		\caption{Cluster 2}
	\end{subfigure}
	\begin{subfigure}{0.3\linewidth}
		\centering
		\includegraphics[width=\linewidth]{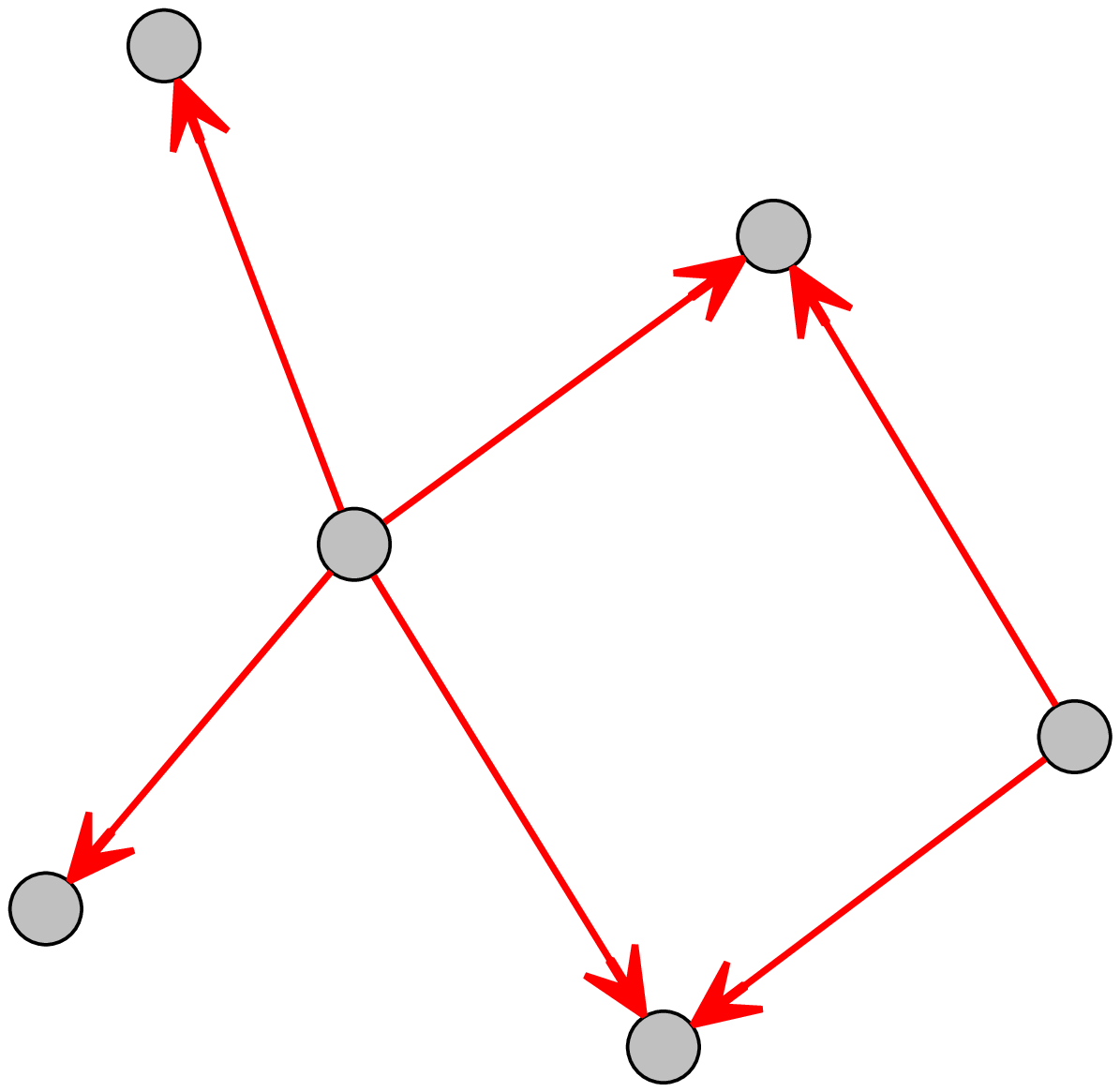}
		\caption{Cluster 3}
	\end{subfigure}
	\begin{subfigure}{0.3\linewidth}
		\centering
		\includegraphics[width=\linewidth]{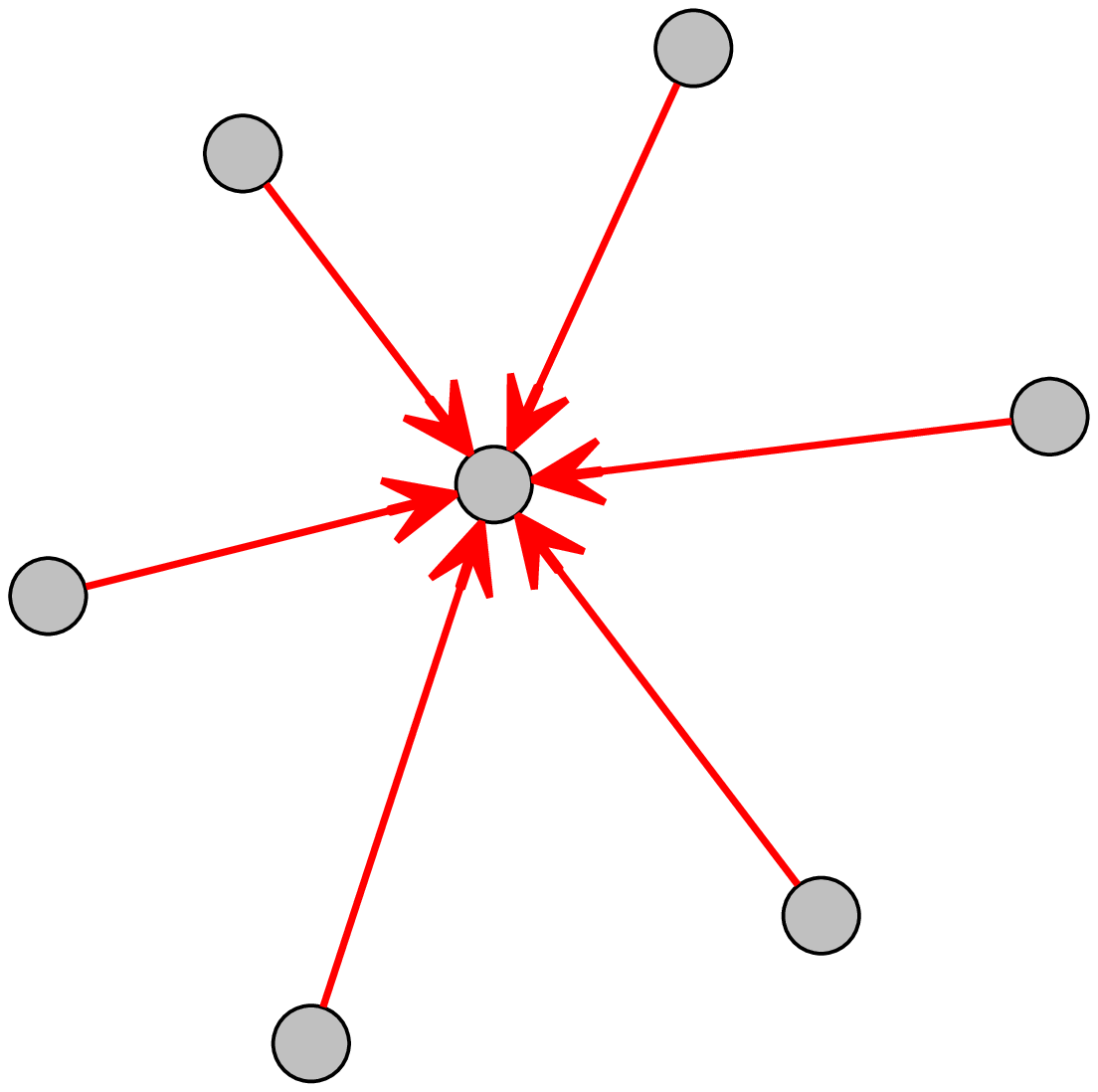}
		\caption{Cluster 4}
	\end{subfigure}
	\begin{subfigure}{0.3\linewidth}
		\centering
		\includegraphics[width=\linewidth]{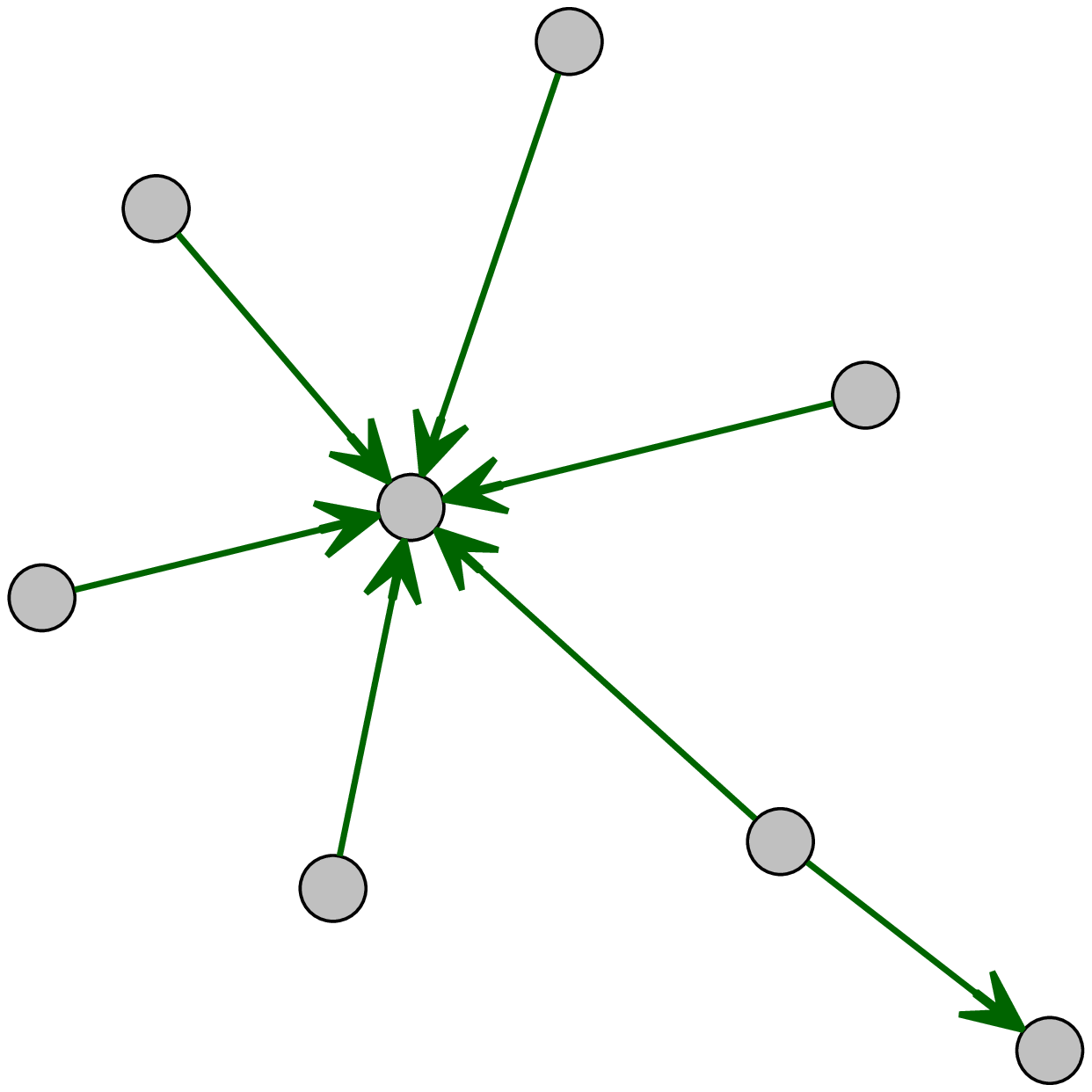}
		\caption{Cluster 5}
	\end{subfigure}
	\caption{
		Example aggregated graphs of representative temporal components from each cluster.
    Messages and retweets are coloured green and red respectively.
    In cluster 1, messages are sent back and forth between multiple nodes, typical of what one would associate with a \emph{conversation}.
		Cluster 3 is associated with one or more central nodes being retweeted by a set of other nodes.
		Cluster 4 consists solely of a single node retweeting many other nodes, and cluster 2 exhibits similar mass retweeting behaviour combined with a small fraction of messaging, usually across a wider array of nodes.
		Finally, in cluster 5 many nodes message one or more central nodes.
  }
	\label{fig:examples}
\end{figure}

In Table~\ref{table:component_statistics} we report the average values for the features within each cluster, combining the motifs with different event types for brevity.
In Figure~\ref{fig:examples} we pick a representative component from each cluster and plot the static aggregated graph of connections between nodes.
Each cluster is distinct and captures some of the different types of behaviour observed on Twitter.

\textbf{Cluster 1.} Cluster 1 boasts the highest edge density, reciprocity, and clustering coefficient.
This cluster is therefore a candidate for what would be classified as \emph{conversational behaviour}; users sharing messages back and forth between each other and with conversations lasting around nine minutes.
Cluster 1 also has the smallest number of nodes on average.

\textbf{Cluster 2.} Cluster 2 consists of hybrid retweeting and message behaviour. 
Nodes will retweet multiple sources, but also send out messages (sometimes to accounts they have retweeted).
This hybrid behaviour often leads to these hubs being connected to other hubs which results in a larger number of connected nodes.
The hub nodes in these components appear to be maximising their activity on the network and exposure to other nodes, however unlike in cluster 1, this interaction is not reciprocated.

\begin{figure*}[!h]
  \centering
  \includegraphics[width=0.8\linewidth]{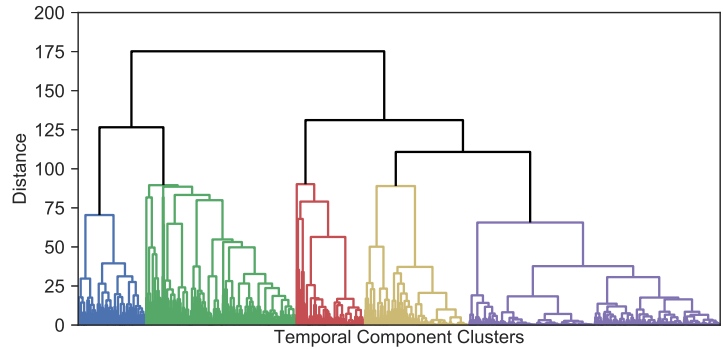}
  \caption{
  Dendrogram of temporal components and associated clustering.
  Pairwise Euclidean distances are calculated and a hierarchy is created using Ward's method.
  The number of clusters is calculated by maximising the silhouette score (see text).}
  \label{fig:dendrogram}
\end{figure*}

\textbf{Cluster 3.} Cluster 3 has an exceedingly large activity rate, compared to the other clusters.
This is due to it containing conversations where many users retweet one or more central bodies.
Retweeting is an instantaneous action which requires little thought and so large numbers of retweets can happen in a short period of time.
This cluster is also short-lived however, lasting less than three minutes on average.
Another observation we can make is the presence of `bridge' nodes which have retweeted two or more of the central nodes.
These users are potentially bringing together two different conversational topics or merely collecting tweets surrounding a single topic.

\textbf{Cluster 4.} Cluster 4 is an example of hub-like behaviour, indicated by the extremal values of $\mu$. 
A single node will retweet many other nodes with little to no other interaction.
The second most prevalent motif in this cluster is ABrABr which indicates the central node is retweeting the same node multiple times.
In this cluster there is little evidence of social interaction as components are primarily driven by a single node aggregating information from other nodes.

\textbf{Cluster 5.} Cluster 5 is in some sense the opposite to cluster 4.
Multiple nodes message a central node, however there is little interaction in the opposite direction, or any sign of diverse behaviour (given the second lowest motif entropy).
These components last substantially longer than their cluster 4 counterparts, despite having fewer events on average (751 seconds compared to 445 seconds).
This is likely due to messages taking longer to compose than an instant retweet, naturally elongating the time over which we see the behaviour.

\subsection{Evolution over time}

\begin{figure}[!h]
	\centering
	\includegraphics[width=0.8\linewidth]{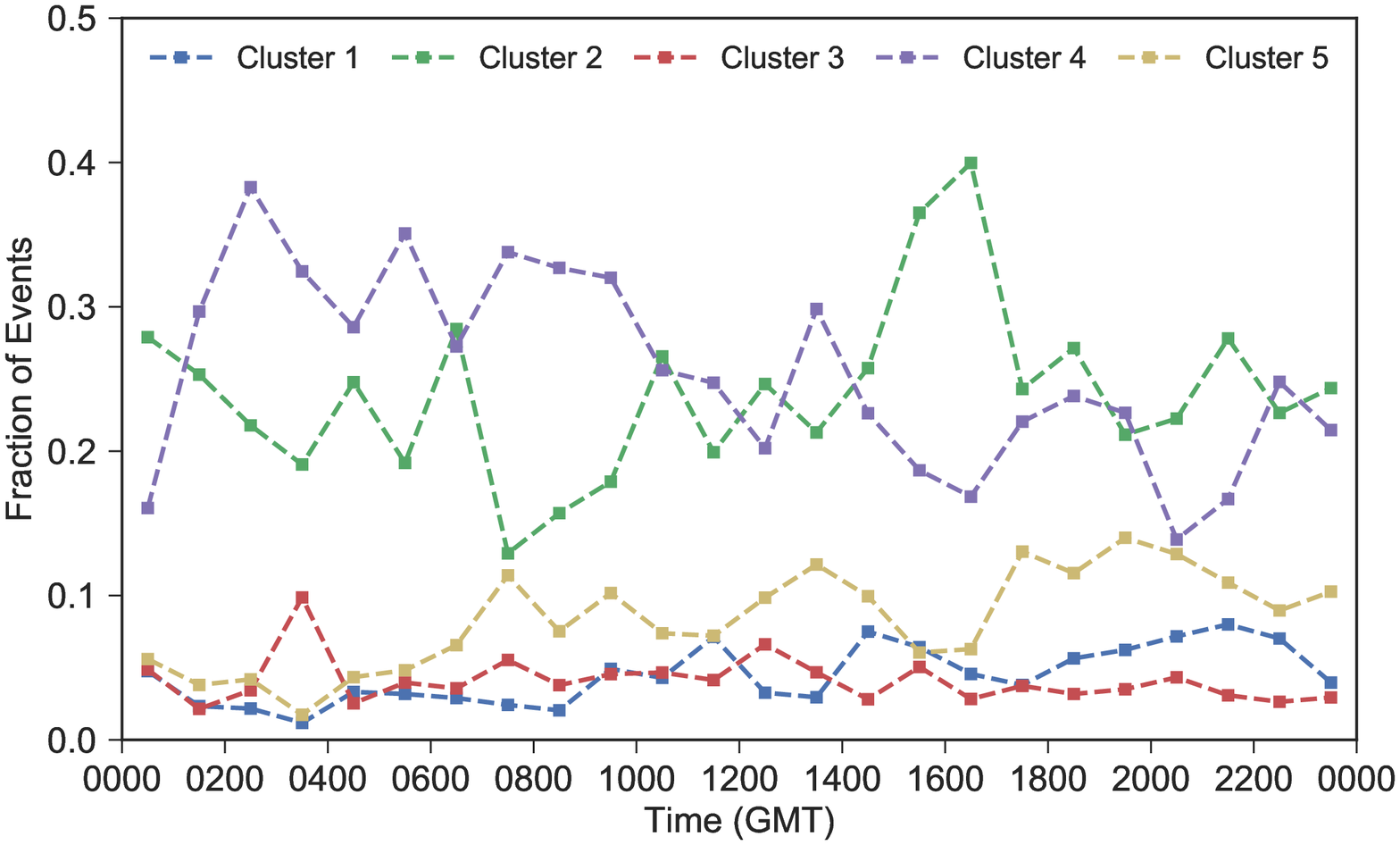}
	\caption{
		Cluster volumes over time as a fraction of all events, aggregated by hour.
		Clusters 2 and 4 are consistently dominant throughout the day, comprising of over half of all events in total.
		The other clusters are less prevalent although their activity levels change over time.
		For example, cluster 5 becomes more prevalent between the hours of 7am and midnight (GMT).
		The remaining fraction of all events are those components with fewer than five events which were not part of the clustering.}
	\label{fig:evolution}
\end{figure}

These prevalence of these clusters are not necessarily equal in size nor do they persist uniformly across time.
Each temporal event is associated with a temporal component which is in turn associated with a cluster.
In Figure~\ref{fig:evolution} we plot the size of a cluster over time as a fraction of all events, aggregated by hour.

Clusters 2 and 4 make up the majority of observed behaviour on the temporal network.
This means that the majority of activity occurring is either single-node retweeting aggregation or a hybrid of aggregation and unreciprocated messaging.
This suggests that the majority of activity on the network is driven by a small number of nodes which may occasionally interact with each other.
The more conversational cluster 1 never contributes to more than 10\% of activity over the course of the day, although naturally becomes more prevalent during daylight hours (for Europe and the Middle East).

\subsection{Comparison with Random Ensembles}

The results of the clustering have highlighted the diverse behaviour on Twitter, however we still need to assess whether this diversity is significant.
To do this we consider an ensemble of 200 time-shuffled samples of the data.

For each sample we first create an `complete' feature vector $\bm{x}^*$ for the dataset by considering the entire temporal network as a single temporal component.
Then for each temporal component feature vector $\bm{x}_c$ we subtract off the complete feature vector, i.e.
\begin{align*}
  \bm{\hat{x}}_c' = \bm{\hat{x}}_c - \bm{x}_*.
\end{align*}
The average distance between the each temporal component the entire network, given by
\begin{align*}
  \frac{1}{|C_{\Delta t}|}\sum_{c \in C_{\Delta t}} |\bm{\hat{x}}_c'|,
\end{align*} 
is a measure of the diversity of components within the network. 
For the ensemble of randomised networks, the mean and standard deviation are 0.648 and 0.002 respectively.
In contrast, the mean of $|\bm{\hat{x}}'|$ for the original data is 0.70 (which comparing to the ensemble distribution equates to a $z$-score of 17.4).
Hence we can conclude that the components features that we are measuring are significant, and are more diverse than we would expect to see at random.
This also supports the decomposition of the temporal network; there are different behaviours across the nodes of the temporal network (and over time) so we may lose this diversity by considering the temporal network as a single entity.

\begin{figure}[!h]
  \centering
  \begin{subfigure}{0.49\linewidth}
    \centering
    \includegraphics[width=\linewidth]{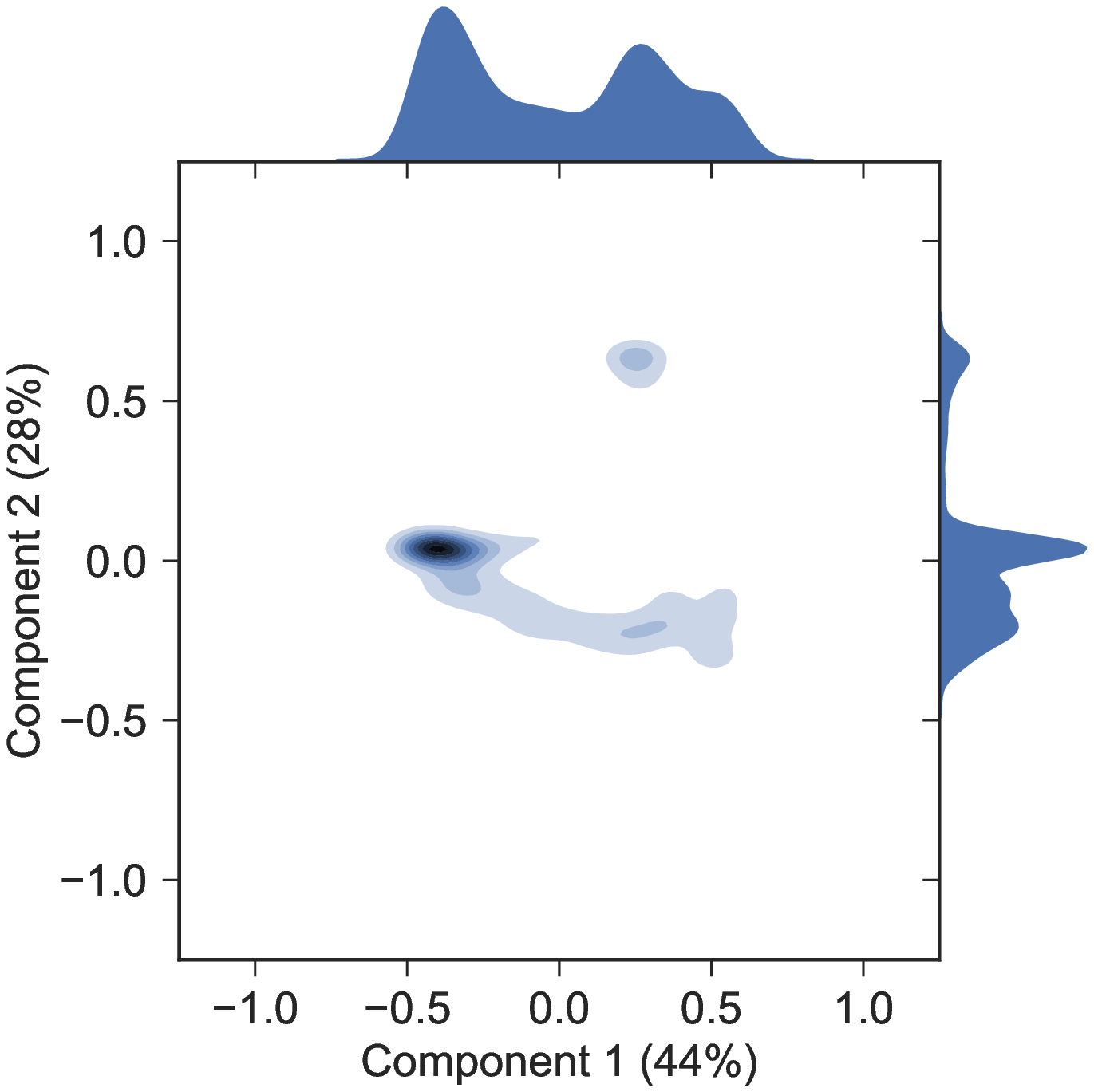}
    \caption{}
  \end{subfigure}
  \begin{subfigure}{0.49\linewidth}
    \centering
    \includegraphics[width=\linewidth]{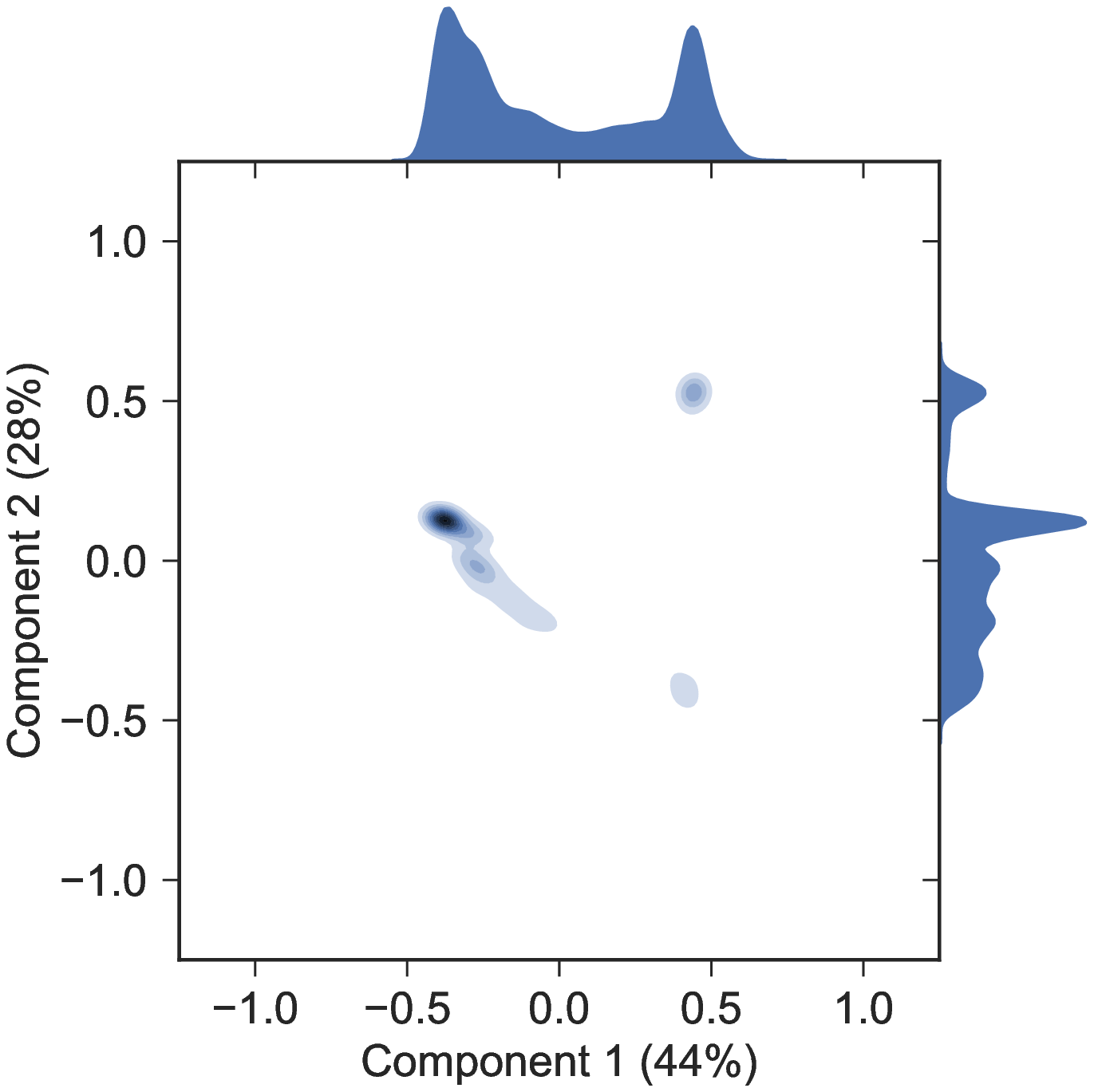}
    \caption{}
  \end{subfigure}
  \caption{
  Temporal component feature densities for the data (a) and an ensemble of 200 time-shuffled instances of the data (b).
  Feature space is reduced to two dimensions using PCA.
  Marginal distributions for each dimension are also shown.
  The original data is more diverse than the time-shuffled ensemble, although the time-shuffled data exhibits some bimodality.
  }
  \label{app:fig:density}
\end{figure}

In Figure~\ref{app:fig:density} we show the density of the temporal components for the original  data (a) and random ensemble (b) in two dimensional feature-space, reduced by principal component analysis. 
The wider marginal distributions in (a) confirm our statistical analysis showing that the temporal components are more diverse than random components.
There do however appear to be a number of possible clusters in the randomised data.
This is likely due to components with repetitive behaviour (such as in cluster 4) where the same motif will appear regardless of the ordering of events.
The primary difference in the shuffled cases is that on average these components will be smaller and there will be more of them.

Finally, as well as a more homogeneous collection of temporal components, there are fewer temporal components (of minimum size five) in the random ensemble than in the original data.
There are on average 3530 temporal components in the random ensembles, compared with 4137 in the original data.
Interestingly, there are more events per component in the original data (21.1 compared to 13.7 on average), however the temporal components in the random ensemble are on average of a longer duration (774s for random ensemble and 536 for original data).
This is likely due to the events for any temporal component being more uniformly spread in contrast to the original data where multiple events may occur within a short period of time (those in cluster 3 for example). 

\subsection{Comparison with Fixed-width Interval Decomposition}

To compare the temporal component decomposition we use a fixed-width interval decomposition of the temporal network.
Following the same procedure outlined above, we calculate the feature vector for each time interval before rescaling and clustering.
We choose the interval width to be the average component duration in the data which is 536 seconds.
The interval-based clustering is poor however, taking a maximal silhouette coefficient of less than 0.2 for two clusters which decreases with the number of clusters (see Figure~\ref{app:sec:clusters}).

\begin{figure}[!h]
	\centering
	\begin{subfigure}{0.49\linewidth}
		\centering
		\includegraphics[width=\linewidth]{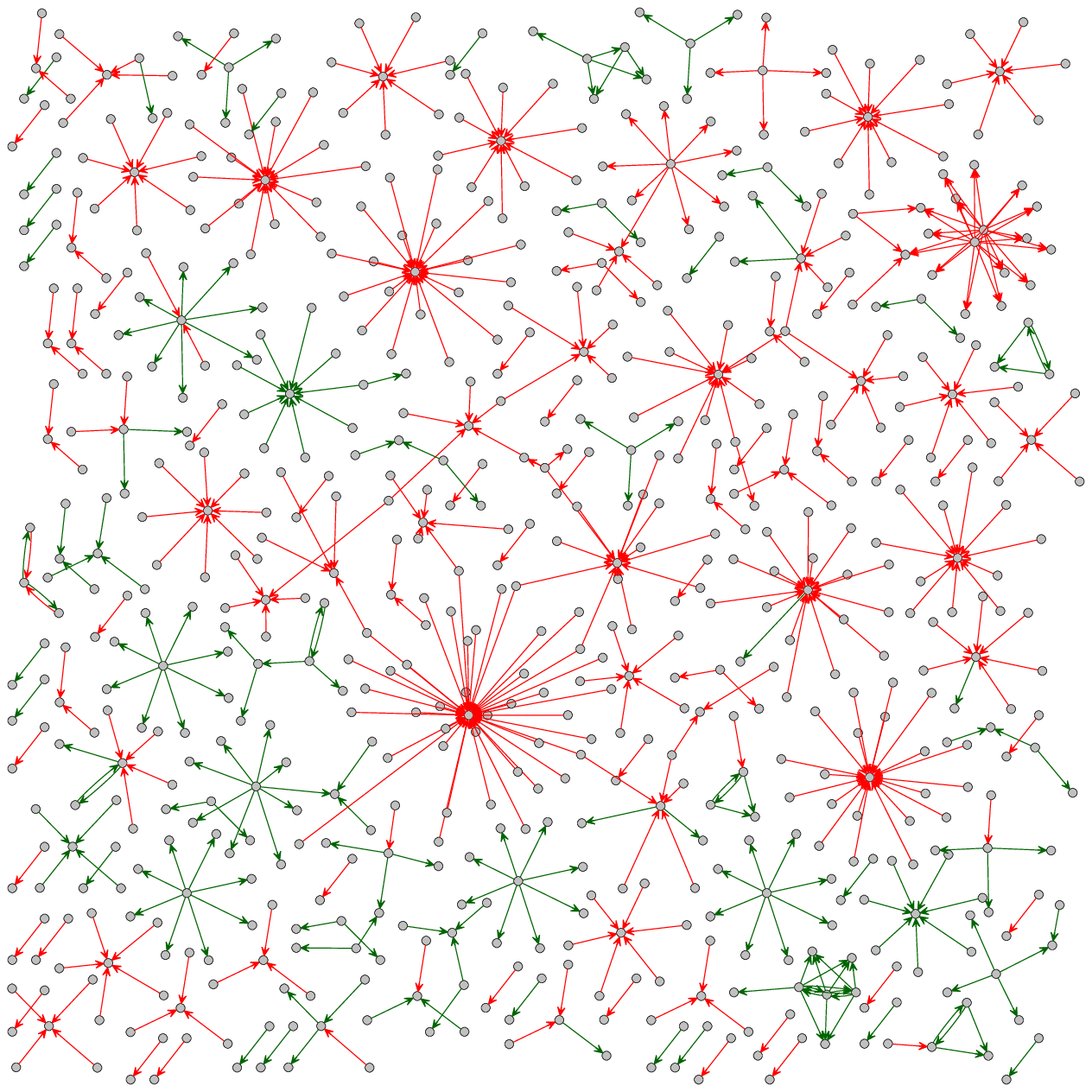}
		\caption{Cluster 1}
	\end{subfigure}
	\begin{subfigure}{0.49\linewidth}
		\centering
		\includegraphics[width=\linewidth]{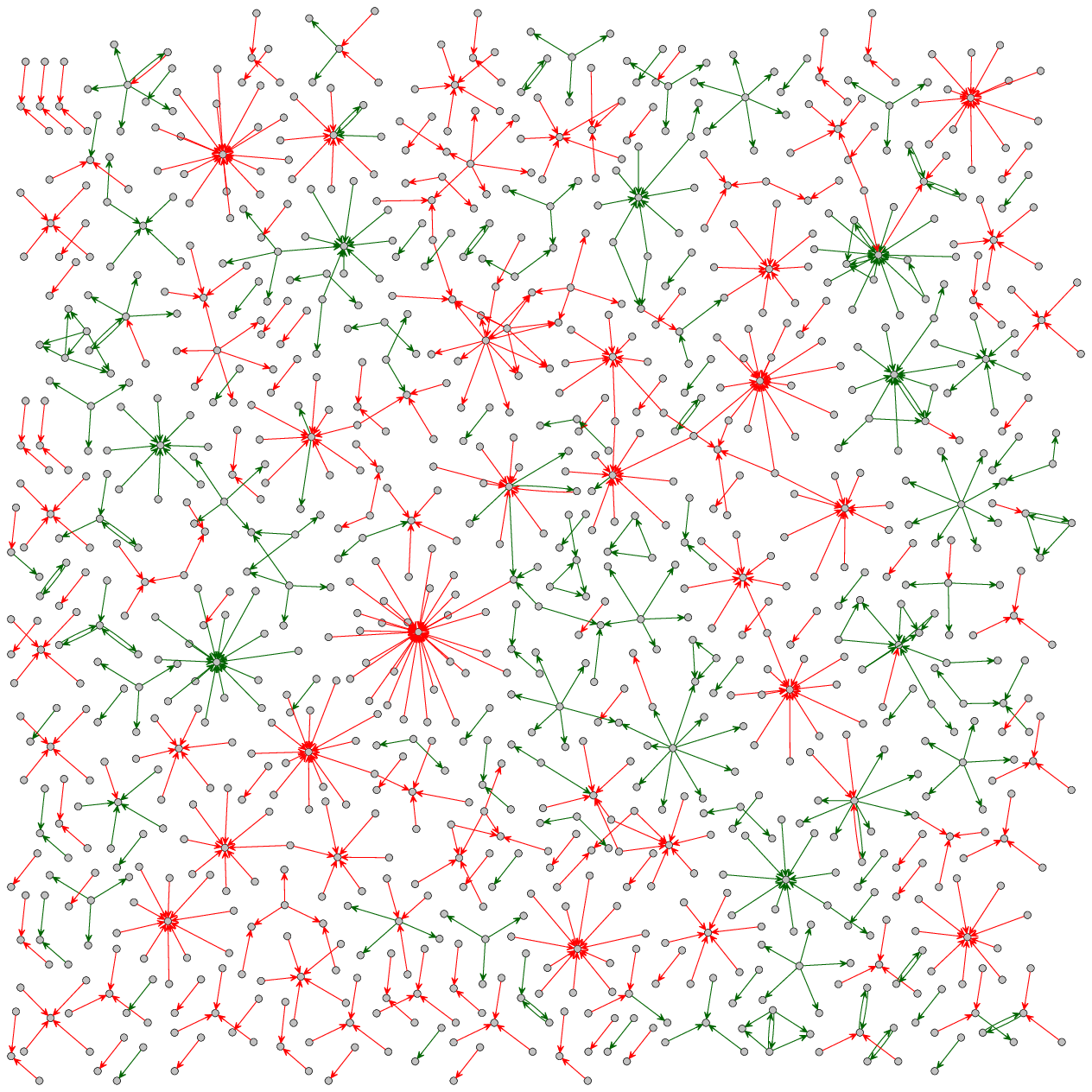}
		\caption{Cluster 2}
	\end{subfigure}
	\caption{
		Example aggregated graphs from the time-interval decomposition.
		The clustering method finds two overlapping clusters which are indistinguishable by eye.
	}
	\label{fig:interval}
\end{figure}

In Figure~\ref{fig:interval} we show examples of the to most representative intervals for each cluster.
Using this information alone it is difficult to distinguish any particular behaviour as by definition all the behaviour is restricted to a particular interval and will therefore overlap.
Based on this evidence there is no real change in collective behaviour over the day long period.
We can however consider a further decomposition into the static components of each time interval.
This results in 5912 different components which we can cluster, more than the 4137 from the temporal component decomposition.
Clustering the interval-based components gives a better decomposition than clustering the intervals alone, however it falls short of the temporal component decomposition (based on the silhouette coefficient across all possible number of clusters).
Furthermore the interval-based components are restricted to have a duration less than the interval width when a particular pattern of behaviour may in fact span multiple intervals.
This is illustrated by considering how many temporal components span multiple intervals.
For this particular interval width, over 60\% of temporal components span multiple intervals and would therefore be split under such a method. 
This could lead to double-counting components, or their omission altogether if the split falls under the minimum component size threshold.
In contrast for a window size of 4 minutes over 80\% of temporal components are split, and for a window size of 20 minutes 37\% of components are split.

\section{Conclusion and Discussion}
\label{sec:conclusions}

In this article we have devised a novel way to decompose a temporal networks using both topological and temporal features.
Furthermore, we combined temporal motifs, inter-event times, and a number of static graph features to create an embedding of the temporal components which can be used to cluster collective behaviour.
We applied this method to a dataset of digital communication taken from Twitter to highlight the diversity of communicative behaviour, however this analysis can be easily extended to other forms of digital communication (email etc.), clickstreams, as well as to other contexts such as biological protein networks or temporal networks of proximity.

There is currently not a suitable benchmark for the evaluation of temporal network embeddings, due to the limited work in this area.
Furthermore, as an original object of study, the temporal components of the event graph have no direct comparison in the literature.
We anticipate that the temporal network embedding and the use of temporal components will need to be validated independently.

The use of each feature should be fully justified.
In principle, the features which are not purely temporal could be captured by considering higher-order motifs.
Here we have used features which capture some of these higher-order interactions (such as clustering) at a cost of approximating the temporal network by a static network.
Since we are considering temporal components of small timescales (minutes), relative to the time period (1 day), this approximation is reasonable.
For future work it would be enlightening to compare the trade off between computation required and the quality of clustering as we consider a balance of higher-order motifs and static approximations.

One feature of this analysis that has not previously mentioned is that it can be conducted in near-real-time.
This is crucial if one wants analyse event streams on timescale of minutes rather than days.
This is particularly useful for Twitter where conversations are typically short-lived.
The computational cost of building the event graph scales linearly with the number of events, and the features of each temporal component can then be computed in parallel.
As these components are small relative to the entire network their summary statistics can be computed quickly.
This allows us to monitor event streams over time and be reactive to change. 

Finally, we discuss the particular merits and issues within the example context.
Developing a greater understanding of online conversations can help both companies and scientists alike understand how we behave in an online social context, and how this behaviour varies over time.
This goes beyond the simple individual statistics (message counts, number of followers etc.) that are commonplace in this area.
Characterising this behaviour allows companies to tailor their activity to the type of conversation occurring and target resources where they will be most effective, potentially leading to more efficient and impactful social media campaigns and interactions.
One particular context where these methods may be useful is in the identification of automated of `bot' accounts.
Current state-of-the-art methods use a number of temporal, network, and language derived features \cite{davis2016botornot}, however they do not consider behaviour from the temporal-topological perspective of temporal motifs (or the associated entropy).
In preparing this work we have also found instances where individual accounts show no abnormal behaviour, but operate systematically as a collective. 
As automation becomes more advanced, so too will be the tools required to detect these accounts.

In this article we have considered only the network structure and have omitted any other associated data (such as text, photos, videos, sentiment).
One possible extension would be to augment this analysis with natural language processing techniques and topical analysis to further understand what is being discussed within conversations, and whether the temporal components lie along topical boundaries.
The simplest addition to this analysis would be calculating the sentiment \cite{pang2008opinion,pak2010twitter} of each conversation.
Sentiment is unreliable for short messages and so the average sentiment of a conversation may be more enlightening than individual messages.
Another point to consider is that this analysis is on a relatively small dataset and the results should not be generalised, the intent of this article being to showcase the method. 
We can apply this analysis across longer time periods and different types of data (such as email) to gain a greater understanding of the diversity of the our digital conversations and temporal networks in general.

\section*{Funding}

This work is supported by the Oxford-Emirates Data Science Lab (OEDSL).

\section*{Acknowledgements}

Thanks to Roxana Pamfil, Madhurima Sinha, Peter Grindrod, and Renaud Lambiotte for useful comments and discussion.

\section*{Availability of data and material}

\begin{sloppypar}
The dataset, code, and notebook supporting this article are available under the folder \emph{/examples/collective\_behaviour\_paper} at \href{https://github.com/empiricalstateofmind/eventgraphs}{https://github.com/empiricalstateofmind/eventgraphs}.
\end{sloppypar}

\newpage
\appendix

\section{Methods}

In this section we give further details on the methods used in this analysis.

\subsection{Event Graph Decomposition}
\label{app:sec:eg_algorithm}

Below we outline the steps to construct the event graph and perform a temporal decomposition by thresholding the edge weights.
This algorithm is implemented in the \emph{eventgraphs}\footnote{
  \url{https://github.com/empiricalstateofmind/eventgraphs}
} Python package, available freely online.

Beginning with an empty graph:
\begin{enumerate}
  \item For each node $x$ construct a time-ordered sequence of events $S_x = (e_k)_{k=1}^{|S_x|}$ such that $e_i \in S_x$ if and only if $x \in \{u_i,v_i\}$, i.e. the set of events for which $x$ is a participant.  
  \item For each consecutive pair of events $e_{k}, e_{k+1}$ in $S_x$, add an edge from $e_k$ to $e_{k+1}$ in the event graph with weight $t_{k+1} - t_{k}$.
  Repeat this process for each node in the temporal network.
\end{enumerate}
The edges of the event graph can then be thresholded to removed all edges over a set value $\Delta t$.
The weakly-connected components of the event graph can then be found through standard methods by considering the corresponding undirected network. 

The event graph can also be constructed in real-time to address streams of event data.
This can be achieved by maintaining a list of the last events that each node has participated in.
Let $\omega_x$ be the event which node $x$ last participated.
Upon arrival of a new event $e_* = (u_*, v_*, t_*)$, 
For each node $x \in \{u_*,v_*\}$, create an edge from $\omega_x \to e_*$ (if $\omega_x$ exists, otherwise no edge is created), and then set $\omega_x = e_*$.
The arrival of a new edge can be processed by considering two lookups and is therefore an inexpensive $\mathcal{O}(1)$ operation.

\subsection{Temporal Motifs}
\label{app:sec:temporal_motifs}

\begin{figure}[!h]
  \centering
  \includegraphics[width=0.12\linewidth]{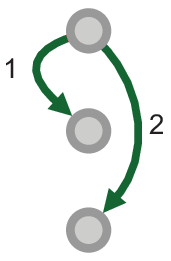}
  \includegraphics[width=0.12\linewidth]{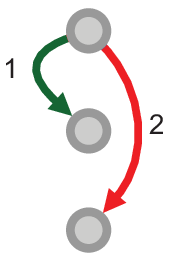}
  \includegraphics[width=0.12\linewidth]{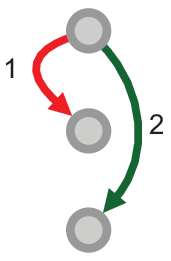}
  \includegraphics[width=0.12\linewidth]{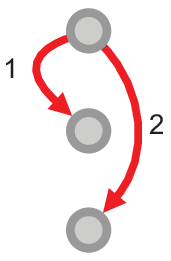}
  \caption{The four possible colourings of the ABAC motif. From left to right these are ABmACm, ABmACr, ABrACm, ABrACr.}
  \label{app:fig:motifs}
\end{figure}

In our feature analysis we consider \emph{coloured motifs}, where edges or events can be of different types, or colours.
We can also consider coloured nodes (to distinguish between node types), however for this study we have no means to distinguish between nodes.
This increases the number of observable motifs by a factor $c^n$ where $c$ is the number of event colours and $n$ is the number of events in the motif.
In Figure~\ref{app:fig:motifs} we show the four possible variations of the ABAC motif.
We adopt the notation ABxACy where x and y are the motif types to describe each motif.

Using coloured motifs allows us to observe more diverse behaviour that would not be captured by simply looking at the counts of each event type.
In the context of Twitter, one behaviour that is often observed is a user being retweeted and then subsequently being messaged by the `retweeter'.
This is usually associated with the retweeter agreeing with a previous post, then adding an original contribution to the discussion.

\section{Data}

In this section we justify our parameter choices for the event graph decomposition and the number of clusters.

\subsection{Decomposition}
\label{app:sec:deltat}

The decomposition of the temporal network using the event graph requires us to choose a value of the parameter $\Delta t$ to threshold the time between to adjacent events occurring for us to consider them to be connected.
There has been little investigation into how the event graph changes with $\Delta t$ \cite{mellor2017temporal} and studies instead consider the sensitivity of final results to variation in the parameter.

\begin{figure}[!h]
	\centering
	\includegraphics[width=0.8\linewidth]{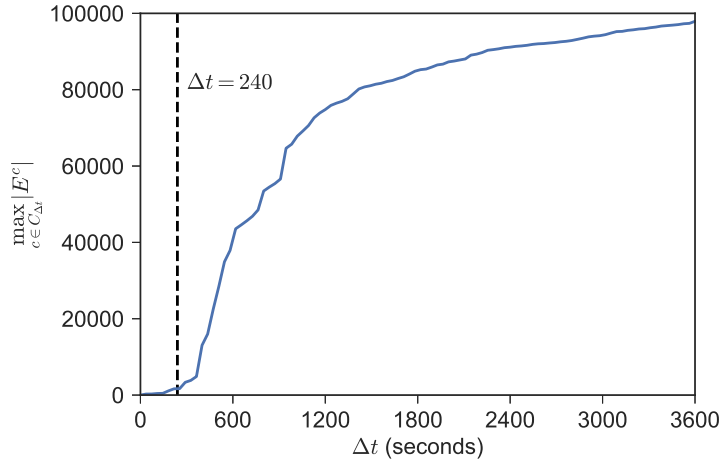}
	\caption{
		Largest temporal size (number of events) as a function of the parameter $\Delta t$.
		The largest component experiences an abrupt transition at approximately 360 seconds, after which the largest component makes up a significant fraction of the entire network (up to 75\% when $\Delta t=3600$).
		We therefore chose a value of $\Delta t=240$ so that the component analysis is not dominated by a single giant component.
	}
	\label{app:fig:deltat}
\end{figure}

In Figure~\ref{app:fig:deltat} we show the size (in terms of number of events) of the largest component of the event graph as a function of $\Delta t$.
We can see that there is an an abrupt growth of the largest component at approximately 360 seconds where the largest temporal component then comprises of a significant number of the events in the network.
At $\Delta t = 3600s$, or one hour, the largest component makes up a majority of all events (approximately 75\%).
As $\Delta t \to \infty$ the number components of the event graph converge to the number of connected components of the aggregate static graph of the entire temporal network, which in this case is 3625 (329 of them have five events or more).  
We therefore choose $\Delta t$ to lie in the range $[0,360]$ so that the number of temporal components remains high.
Testing within the range around $\Delta t = 240$ we find that the clusters found are stable.

Alternatively we could have made a context specific choice for the parameter $\Delta t$, informed by typical response times on Twitter.

\subsection{Number of Clusters}
\label{app:sec:clusters}

Since we are dealing with an unsupervised clustering, we have no measure of clustering accuracy.
Instead we consider the silhouette coefficient (outlined in the main text) which captures the  distance of a sample from the centre of its assigned cluster relative to the centre of the nearest neighbouring cluster.

\begin{figure}[!h]
	\centering
	\begin{subfigure}{0.49\linewidth}
		\centering
		\includegraphics[width=\linewidth]{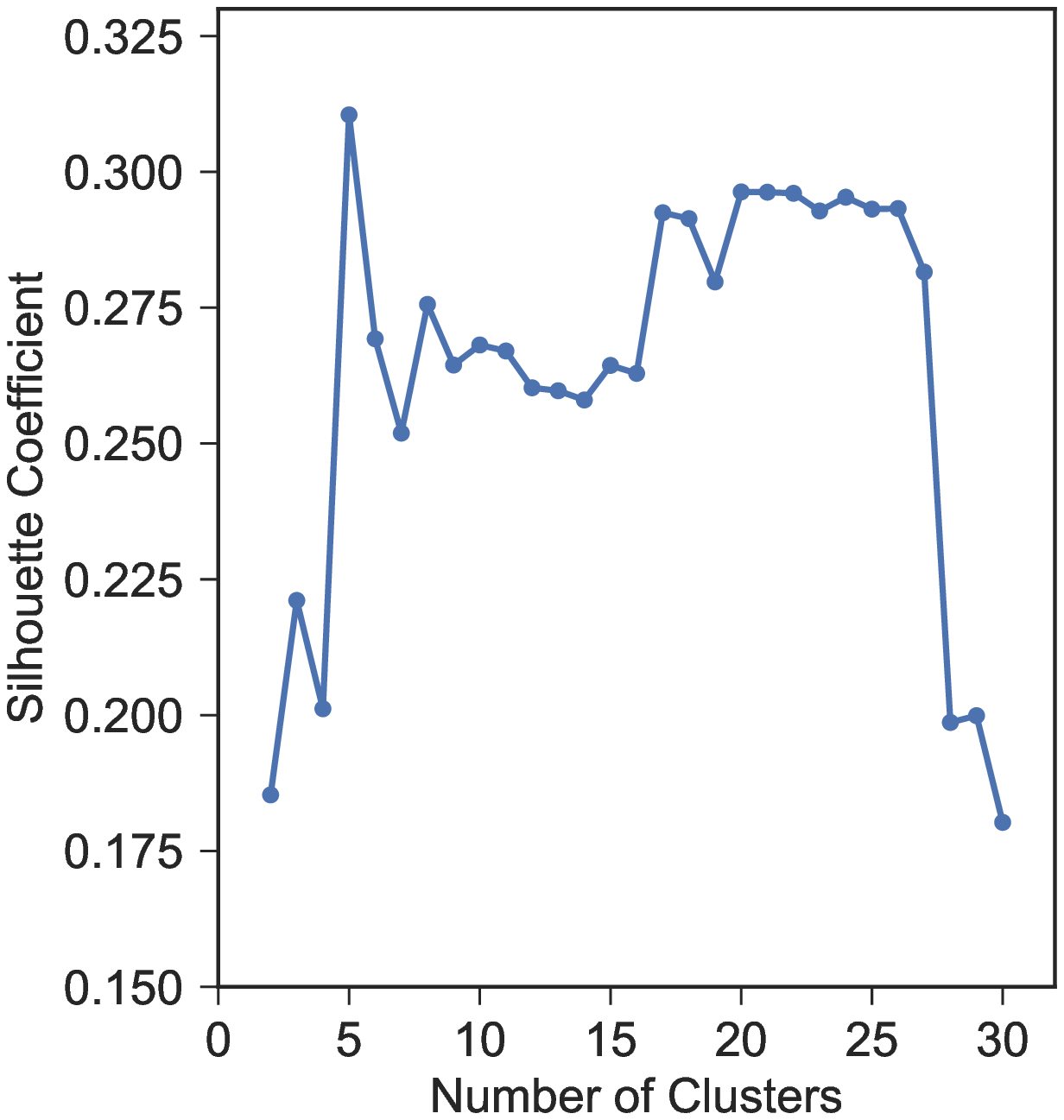}
		\caption{}
	\end{subfigure}
	\begin{subfigure}{0.49\linewidth}
		\centering
		\includegraphics[width=\linewidth]{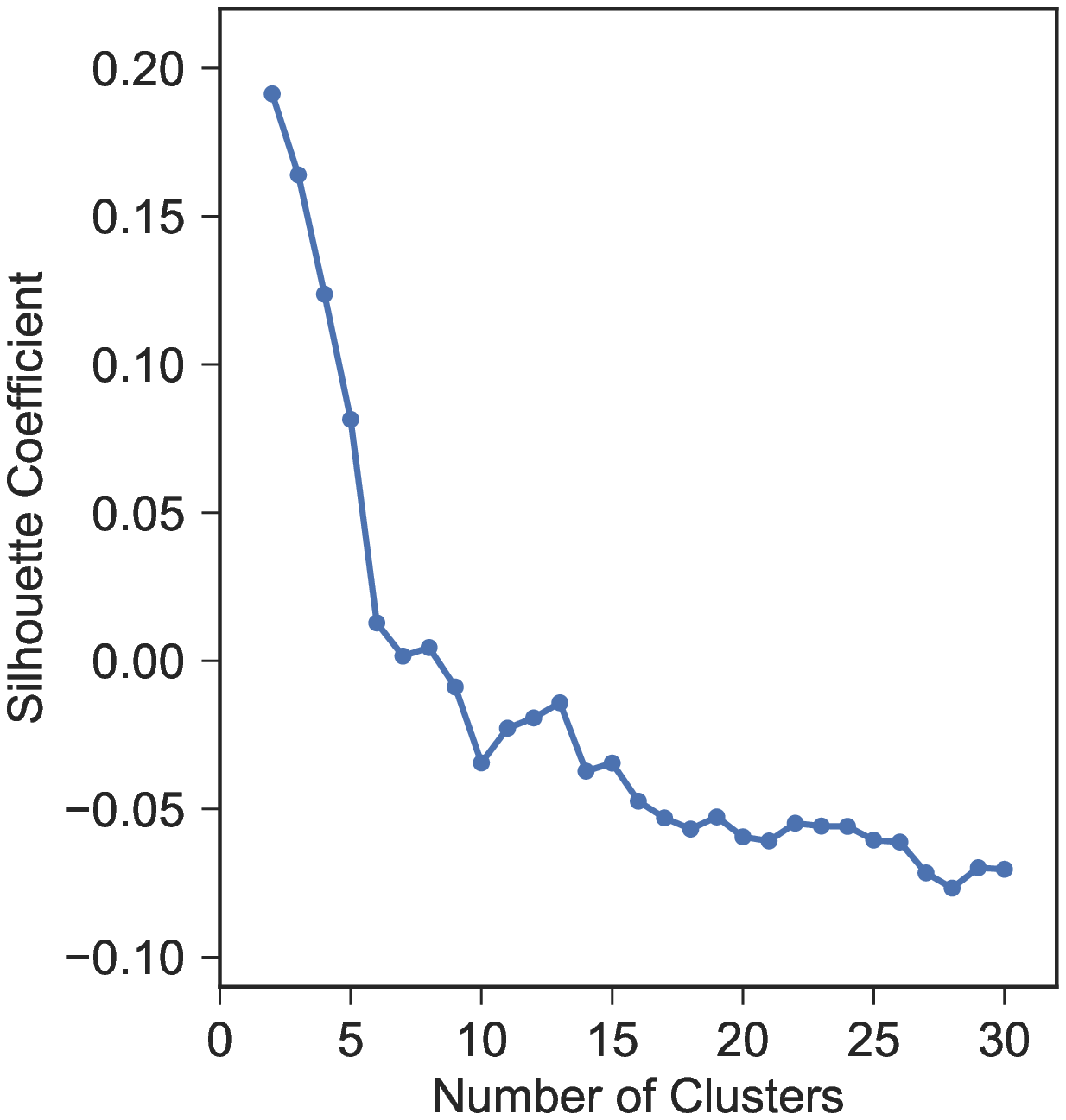}
		\caption{}
	\end{subfigure}
	\caption{
		Silhouette coefficients of the hierarchical clustering for the temporal component decomposition (a) and the interval decomposition (b).
		For the temporal component decomposition the silhouette coefficient is maximal for five clusters, although a drastic drop in cluster performance occurs only at 28 clusters.
		For the interval decomposition, the silhouette coefficient is lower than most values in (a) and roughly decreases with the number of clusters.
	}
	\label{app:fig:clusters}
\end{figure}

In Figure~\ref{app:fig:clusters} we show the silhouette coefficient for the temporal component decomposition (a) and the decomposition into intervals (b).
For the temporal components the clustering is best for five clusters, although the coefficient is comparable for anywhere between 17 to 27 clusters, before seeing a sharp drop in quality for 28 clusters.
This suggests that should more granularity be required (less variance within clusters) then a larger number of clusters could be chosen.
In this case five clusters were chosen to simplify the presentation of the method.

The interval decomposition has a different profile, showing a decreasing trend as the number of clusters increases.
This suggests the best clustering is using only two clusters, however this clustering is relatively poor in comparison to the temporal component decomposition.
With five or more clusters the samples are on average closer to neighbouring clusters than their assigned clusters.

\end{document}